\documentclass{PoS}

\usepackage{outlines}
\usepackage{tabularx}
\usepackage{float}
\usepackage[caption = false]{subfig}
\usepackage{wrapfig}

\title{Estimate of the energy spectrum of the light component of
cosmic rays in HAWC using the shower age and the fraction of 
hit PMT's}

\ShortTitle{HAWC measurements on the cosmic ray spectrum}

\author{
\speaker{J.C. Arteaga-Vel\'azquez}$^{1}$, Z. Hampel-Arias$^{2}$
and J.D. \'Alvarez$^{1}$  
for the HAWC Collaboration\footnote{
A complete list of authors is available at http://www.hawc-observatory.org/collaboration/icrc2017.php.
} \\
\llap{$^1$} Instituto de F\'\i sica y Matem\'aticas, Universidad Michoacana, Morelia, Mexico\\
\llap{$^2$} Department of Physics, University of Wisconsin-Madison, Madison, WI, USA\\

E-mail: \email{arteaga@ifm.umich.mx}       
}

\abstract{The High Altitude Water Cherenkov (HAWC) observatory is a 
ground-based air-shower detector designed to study the TeV gamma and cosmic 
ray windows. The observatory is composed of a densely packed array of 
$300$ water Cherenkov tanks, $4.5$ m deep and $7.3$ diameter with $4$ 
photomultipliers (PMT) each, distributed on a $22,000 \, \mbox{m}^2$ 
surface. The instrument registers the number of hit PMT's as well as the  
timing information and the total charge seen by the photomultipliers during 
the shower event. From the analysis of these data, shower observables such as 
the arrival direction, the core position at ground, the age and the primary 
energy are estimated, from which the energy spectrum of cosmic rays and its
composition can be studied. In this work, we will describe the methodologies of 
HAWC cosmic ray analyses, including the Bayesian spectral unfolding 
procedure used to determine the all-particle and the light component
energy spectra of cosmic rays.  We will see that the distribution of the
shower age vs the fraction of hit PMT's contains information about the 
composition of cosmic rays.}

\FullConference{35th International Cosmic Ray Conference - ICRC217-\\
		10-20 July, 2017\\
		Bexco, Busan, Korea}

\begin{document}

 \section{Introduction}
 
  The energy region between $10 \, \mbox{TeV}$ and $1 \, \mbox{PeV}$
  of the cosmic ray spectrum has been poorly explored as it
  is located at the frontier between the direct and indirect 
  measurements. However, it is of great interest due to the 
  possible presence of new structures in the all-particle
  and the elemental spectra of cosmic rays as predicted, for
  example, in \cite{Stanev} or as the recent measurements 
  performed by ARGO-YBJ seem to suggest, which indicate the
  possible presence 
  of a knee around $1 \, \mbox{PeV}$ in the spectrum of the 
  light mass group (H+He) of cosmic rays \cite{Argo}.  
  HAWC can contribute to the study of both the spectrum and 
  composition of cosmic rays in this poorly explored energy 
  interval. HAWC is an extensive air shower (EAS) observatory 
  located at $4100 \, \mbox{m}$ \textit{a.s.l.} 
  ($\sim 640 \, \mbox{g}/\mbox{cm}^2$ of atmospheric depth) 
  on the northern slope of the volcano Sierra Negra in Mexico.
  The instrument is designed to detect $\gamma$ rays in the 
  $100 \, \mbox{GeV}$ to $100 \, \mbox{TeV}$ energy range,
  but its altitude and physical dimensions permit measurements
  of primary hadronic cosmic rays up to PeV energies.
  In this contribution, we will describe
  analyses that are underway to determine 
  the spectra of the all-particle and the light component 
  of cosmic rays using HAWC data.  
  
  \section{Event reconstruction and simulation}
  
  HAWC is composed of $300$ water Cherenkov detectors, $4.5$ m 
 deep and $7.3$ m in diameter, each instrumented with $4$ upward-facing 
 photomultipliers (PMTs). The instrument covers a $22,000 \, \mbox{m}^2$ 
 area and registers the PMT signals and their respective timing 
 information from extensive air-showers \cite{HAWCcrab17}.
 From these data, the event reconstruction procedure 
 allows estimation of various shower properties such as the arrival direction, 
 the core position at ground-level, the lateral shower age parameter (hereafter 
 referred to as age) and the primary energy. 
 
 For a primary particle interacting in the atmosphere, the resulting 
 air shower is simulated using the CORSIKA \cite{Heck:1998vt} package (v740), 
 with FLUKA \cite{Fluka}, and QGSJet-II-03 \cite{qgsjet}
 as the hadronic interaction models.
 Primaries of the eight species measured by the CREAM \cite{cream} flights 
 (H, $^{4}$He, $^{12}$C, $^{16}$O, $^{20}$Ne, $^{24}$Mg, $^{28}$Si, $^{56}$Fe)
 were generated on an $E^{-2}$ differential energy spectrum from 5 GeV $-$ 3 PeV 
 and distributed over a 1 km radius circular throwing area.
 The simulated zenith angle range was $0^{\circ}-70^{\circ}$, azimuthally symmetric, 
 and weighted to a $\sin{\theta}\cos{\theta}$ arrival distribution, where $\theta$ is the
 zenith angle of the shower axis.
 The secondary charged particles interacting with the HAWC detector were simulated 
 using the  GEANT4 \cite{Geant4} package.
 Dedicated HAWC software was used to simulate the detector response over the entire 
 hardware and data analysis chain.
 Simulations for each particle type are re-weighted to the best fits of a 
 broken power-law to data from direct-detection measurements provided by 
 AMS \cite{ams}, CREAM \cite{cream}, and PAMELA \cite{pamela}, constituting our nominal
 composition model.

 The age, $s$, is obtained event-by-event from a fit to the lateral charge 
 distribution measured by the PMT's at the shower plane with an NKG
 function \cite{NKG}:
 \begin{equation}
 f_{ch}(r) = A \cdot (r/r_0)^{s-3} \cdot (1 + r/r_0)^{s-4.5},
 \end{equation} 
 where $r$  is the radial distance to the shower axis, 
 $r_0 = 124.21 \, \mbox{m}$ is the Moliere radius and $A$ is a normalization 
 parameter.  This lateral distribution function gives a good description of 
 gamma-induced EAS \cite{Kelly}, and a reasonable description of hadron-induced showers 
 at energies $E \lesssim 50$ TeV. At higher energies and large core distances,  
 however, deviations appear between the above function and
 the observed lateral distribution of showers produced by cosmic rays. Further
 research to reduce these differences is now ongoing and will serve to 
 improve the preliminary composition analysis that is presented here.

 \begin{figure*}[!tbp]
  \centering
   \subfloat{
             \includegraphics[width=0.33\textwidth]{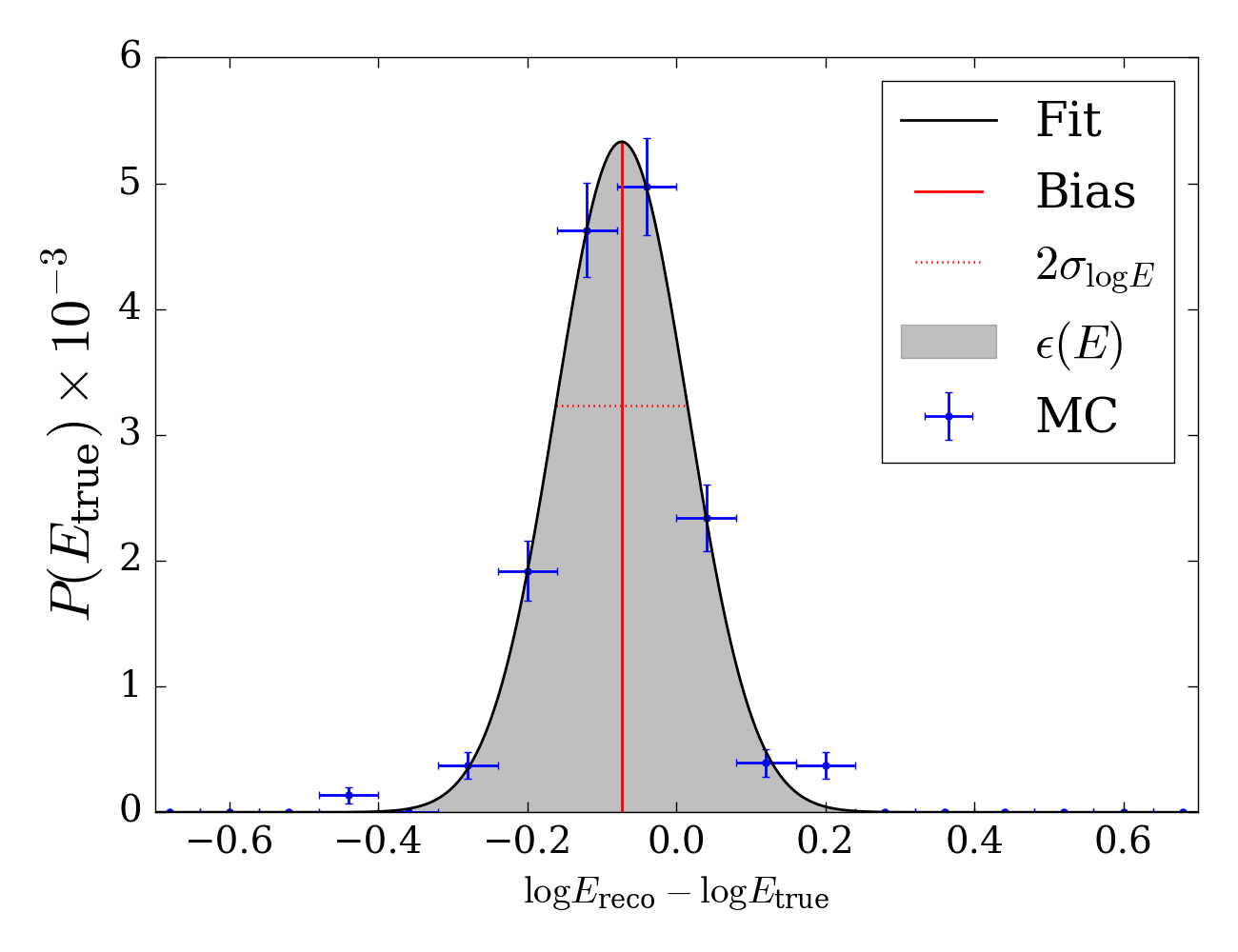}
            }
   \subfloat{
             \includegraphics[width=0.33\textwidth]{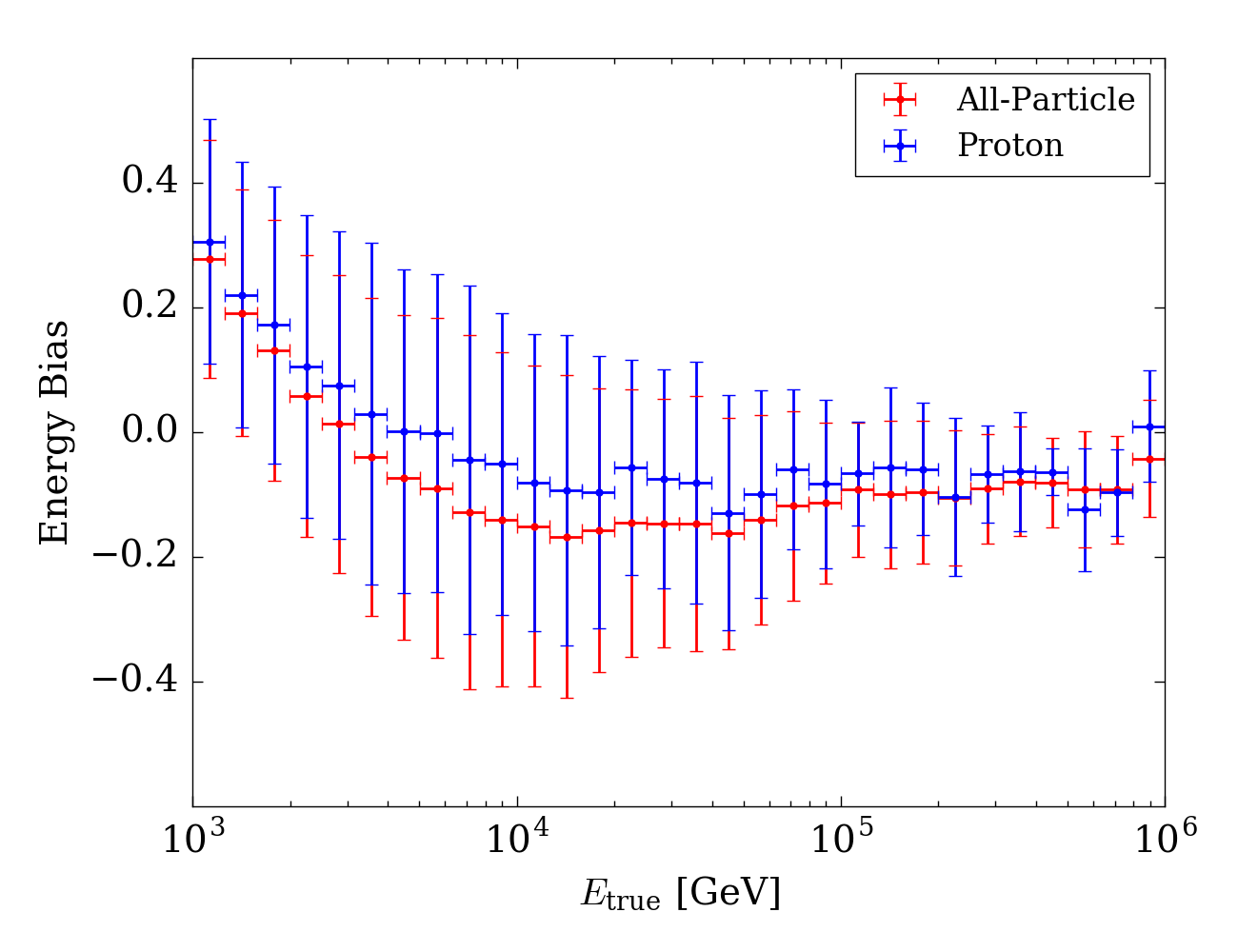}
            }
   \subfloat{
             \includegraphics[width=0.33\textwidth]{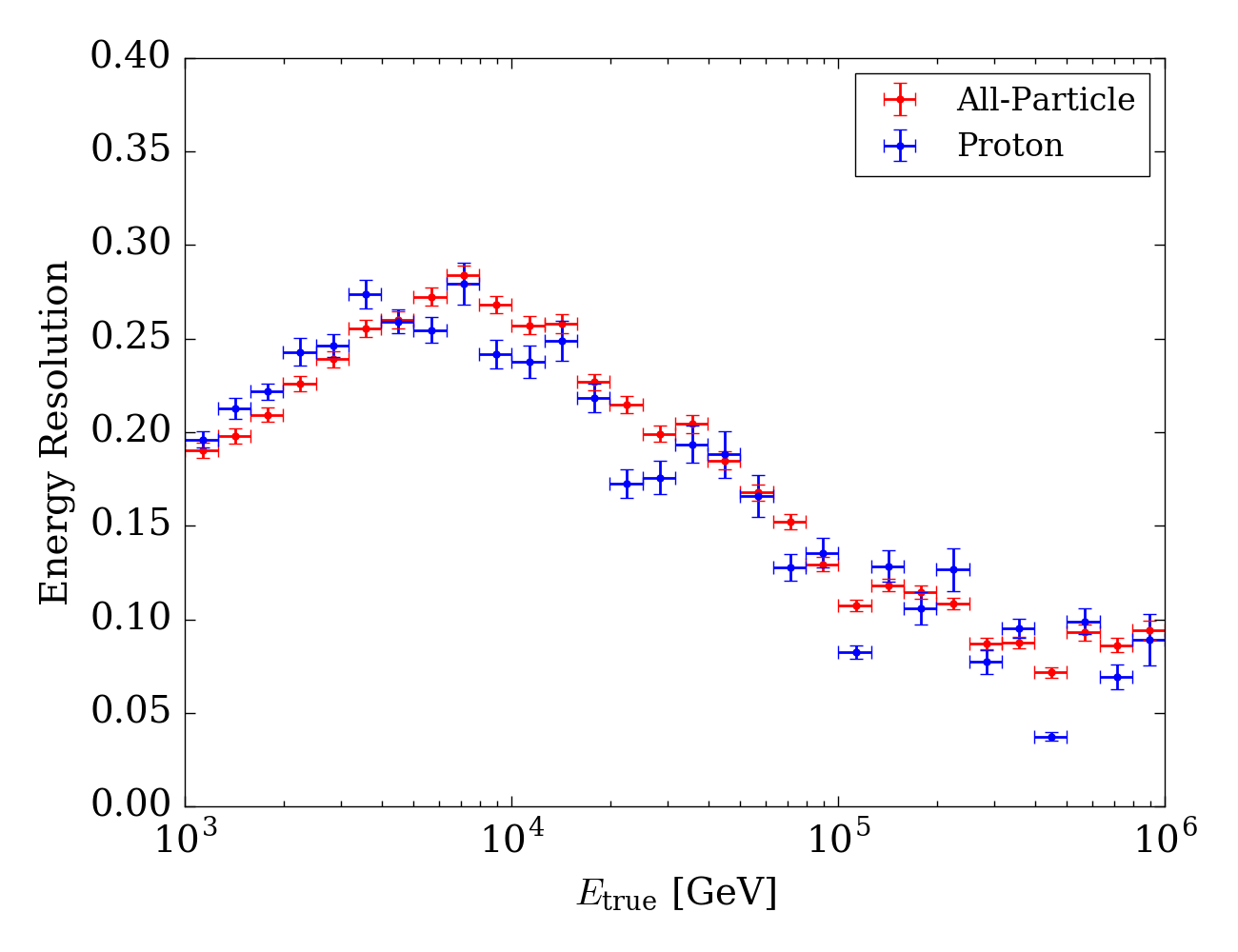}
            }
   \caption{
            Bias distribution for the bin centered at $E_{\textnormal{true}}=100$ TeV 
            (left), showing the definitions of energy bias, resolution, and efficiency.
            The Monte Carlo values are indicated by the blue markers,
            while the black curve is a Gaussian fit to these points.
            The normalization condition includes events that were not selected
            as a result of the selection criteria, thus, the integral represents
            the efficiency $\epsilon(E_{\textnormal{true}})$ 
            to reconstruct and select events in this energy bin.
            Resulting energy bias (center) and resolution (right) for all particles (red) 
            and protons (blue) as a function of the reconstructed energy according to MC data provided the event selection criteria described in the text.
            All panels shown are for the most vertical zenith bin of the energy estimation table.
           }
   \label{fig:lhe-performance}
 \end{figure*}
To estimate the primary cosmic ray energy, we use the lateral 
distribution of measured signal as a function of the primary particle energy.
Using proton-initiated air shower simulations, we build a four-dimensional probability table with bins 
in zenith angle, primary energy, PMT distance from the core in the shower plane, and measured PMT signal amplitude.
Given a shower with a reconstructed arrival direction and core position, 
each PMT contributes a likelihood value extracted from the tables, including operational PMTs 
that do not record a signal.
For each possible energy, the log-likelihood values for all PMTs are summed,
and the energy bin with the maximal likelihood value is chosen as the best energy estimate.

The proton energy table takes the following form:
\begin{itemize}
 \setlength\itemsep{0em}
  \item Three zenith angle bins 
    \begin{itemize}
     \setlength\itemsep{0em}
     \item $\theta_{0}: 0.957 \le \cos \theta \le 1 $
     \item $\theta_{1}: 0.817\le \cos \theta < 0.957$
     \item $\theta_{2}: 0.5 \le \cos \theta < 0.817$
    \end{itemize}
  \item Forty-four energy bins from 70 GeV--1.4 PeV with bin width $0.1$ in $\log E$
  \item Seventy bins in lateral distance from 0--350 m in bins of width $5$ m
  \item Forty bins in charge from 1 -- $10^{6}$ photoelectrons (PE) in steps of $0.15$ in $\log{Q}$
\end{itemize}

The resulting performance is evaluated via the bias distribution,
defined by the difference between the logarithms of the reconstructed and true energy values,
$\textnormal{bias} = \log{E_{\textnormal{reco}}} - \log{E_{\textnormal{true}}}$,
shown for a single true energy bin in the leftmost panel of figure \ref{fig:lhe-performance}.
The mean of this distribution defines the energy bias or offset and the width defines 
the energy resolution, which are shown as a function of energy in the remaining 
panels of figure \ref{fig:lhe-performance}, provided the event selection criteria
for the most vertical zenith bin.
We also identify the integral of the distribution as the efficiency $\epsilon$ 
to reconstruct events in that energy bin.
Above 10 TeV, the bias is estimated to be less than an energy bin width, 
and the energy resolution is $<60\%$, improving with increasing energy.

 \section{The all-particle spectrum from vertical data}
 
 For the all-particle spectrum analysis, we select reconstructed 
 air-showers within the first zenith bin of the energy estimation tables, 
 $\theta < 16.7^\circ$. 
 In addition, we apply a set of selection cuts to reduce the 
 uncertainty on the shower observables. In particular, we choose events 
 that pass the arrival direction and core reconstruction procedures, 
 have a minimum multiplicity threshold $N_{hit} \geq 75$ PMTs, and have activated 
 at least $40$ PMTs within $40$ m of the shower core ($N_{r40} \geq 40$), the 
 latter to ensure EAS with cores on or within $15 \, \mbox{m}$ of the HAWC array.
 Provided the above selection criteria, the core resolution is estimated to be 
 $10$ m at $10$ TeV, dropping below $8$ m above $100$ TeV, while the angular  
 resolution is found to be $\sim 0.5^{\circ}$ above $10$ TeV. 

 \begin{figure*}[!t]
  \centering
   \subfloat{
             \includegraphics[width=0.5\linewidth]{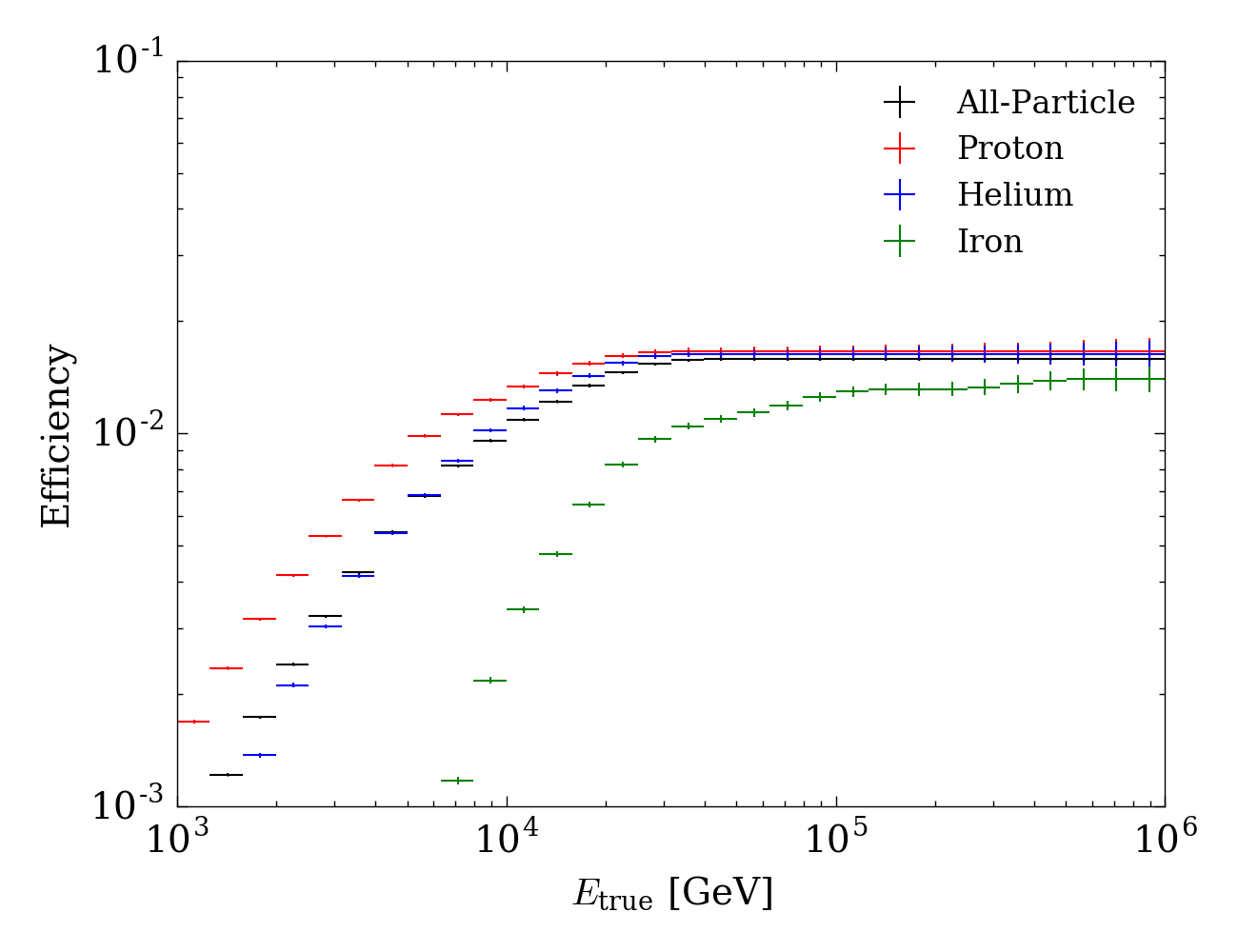}
            }
   \subfloat{
             \includegraphics[width=0.5\linewidth]{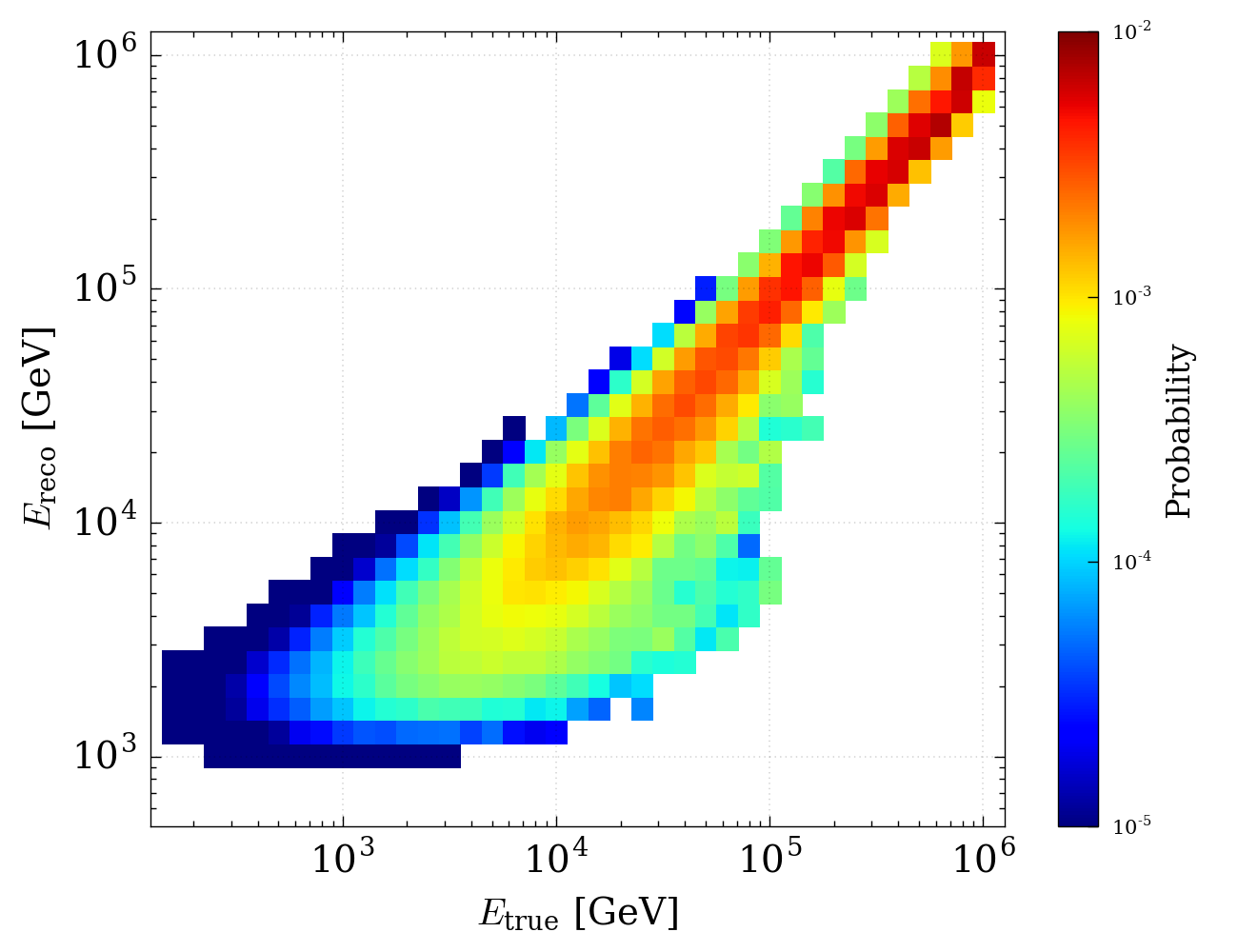}
            }
   \caption{
            The left panel shows the efficiencies $\epsilon(E_{\textnormal{true}})$ 
            for all cosmic ray particles, and proton, helium, and iron components.
            The energy response matrix $P(E_{\textnormal{reco}}|E_{\textnormal{true}})$ 
            for all species using the nominal composition model is shown on the right.
           }
 
   \label{fig:all_part_aeff_mm}
 \end{figure*}
The smearing of the primary particle's estimated energy including detector effects such 
as limited core and angular resolutions must be taken into account in order to measure the 
cosmic ray flux.
This is summarized by the detector response matrix, as shown in figure 
\ref{fig:all_part_aeff_mm} which 
is determined from simulation, including the assumed composition model and selection criteria.
Following the iterative method presented in \cite{agostini}, the observed reconstructed energy 
distribution $N(E_{\textnormal{reco}})$ from data is to be unfolded with the energy response 
to obtain the measured all-particle energy distribution,  $N(E_{\textnormal{true}})$.
The analysis is performed iteratively: an initial estimate of $N(E_{\textnormal{true}})$ is 
convolved with the response matrix providing an improved estimate, which is used again until 
variations on $N(E_{\textnormal{true}})$ from one iteration to the next are negligible.
The differential flux is then calculated from the final $N(E_{\textnormal{true}})$ 
according to the following:
 \begin{equation}
 	\Phi(E) = \frac{N(E_{\textnormal{true}})}{\Delta t 
    \Delta \Omega \ A_{throw} \ \Delta E} \, ,
 	\label{eq1}
 \end{equation}
where $\Delta t$ is the observation time, $\Delta \Omega$ is the solid angle interval, 
$\Delta E$ is the energy bin width centered at $E_{\textnormal{true}}$, 
$A_{throw}$ is the simulated throwing area.
The efficiency, $\epsilon(E_{\textnormal{true}})$, shown in figure \ref{fig:all_part_aeff_mm} 
is taken into account in the unfolding procedure.

     \begin{figure}[!tp]
     \centering
     \footnotesize
     \includegraphics[width=70mm]{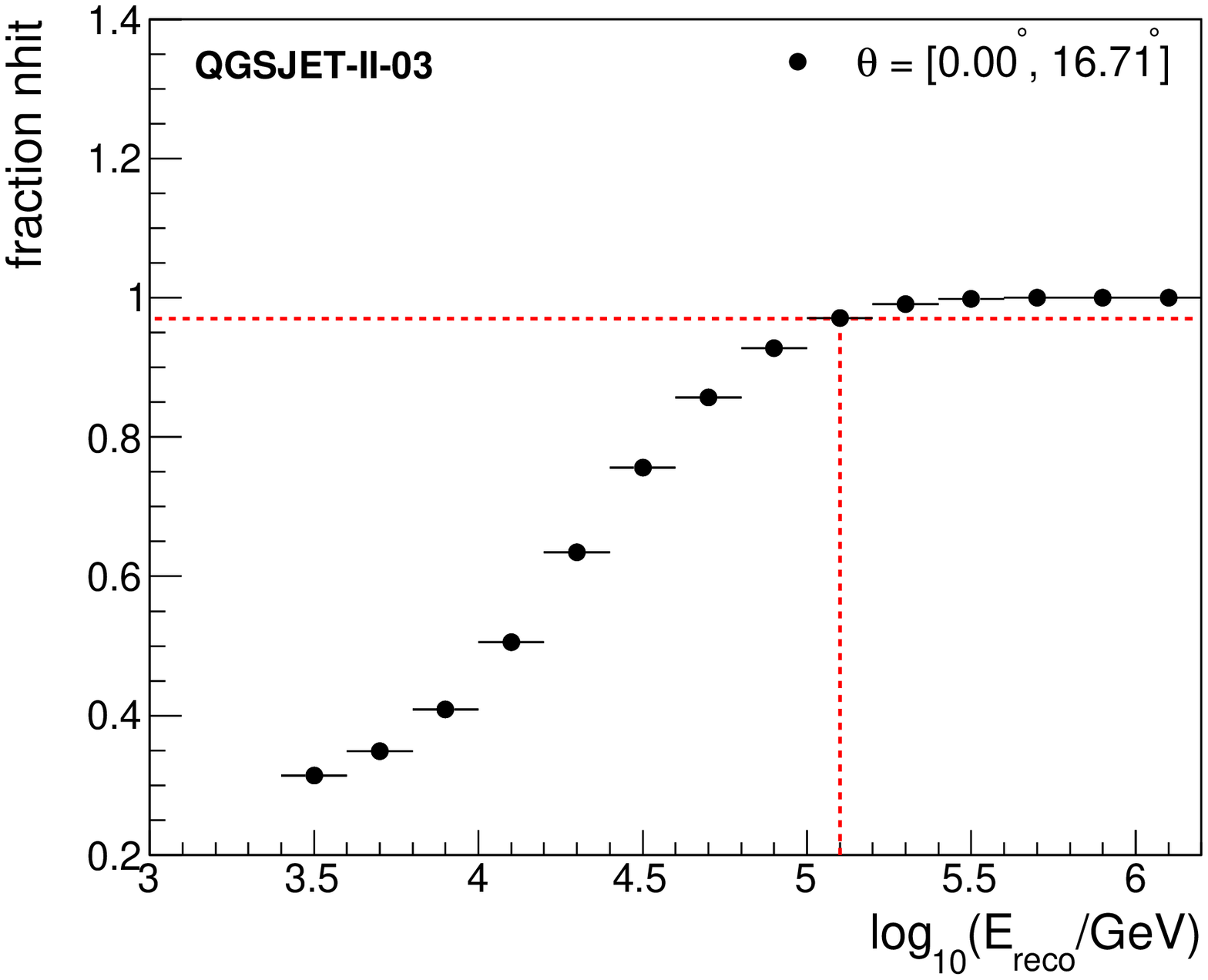} 
     \includegraphics[width=70mm]{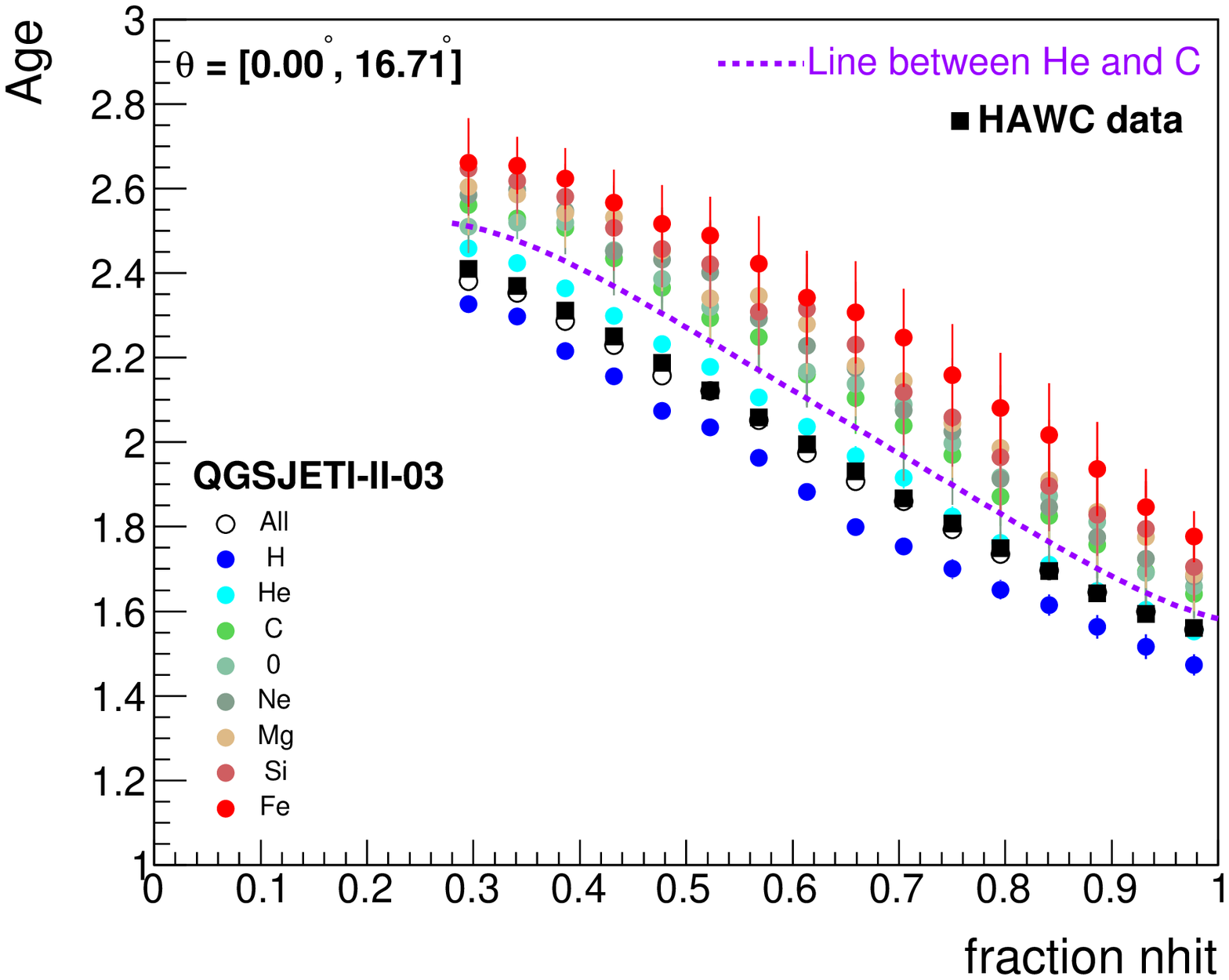}
    \caption{ \textit{Left panel}:  $f_{hit}$ as a function of the reconstructed energy. 
    The horizontal line at $f_{hit} = 0.97$ represents the lower limit of the saturation 
    region and the vertical
    line, the threshold energies where saturation effects begin. Curves were calculated
    using MC data and our nominal composition model.
    \textit{Right panel}: Mean values of the expected age parameter for vertical EAS
    against the estimated
    shower energy for different primaries and for our nominal composition model 
    (labelled as \textit{all}). The HAWC data is also shown for comparison. Errors on 
    the mean are shown. The segmented line represents the $s_{He-C}$ cut.}
  	\label{f1}
\end{figure}

 \section{Spectrum of the light component of cosmic rays from vertical and
 inclined EAS}

  In order to estimate the spectrum of the light component 
 of cosmic rays (H plus He nuclei), we select a subsample 
 dominated by the primaries in which we are interested in. Then we 
 construct its energy distribution, which we again correct for migration 
 effects using an unfolding algorithm. Next, we correct the unfolded 
 histogram for the abundance of heavy nuclei ($Z \geq 3$) and from the 
 result, we reconstruct the corresponding energy spectrum of the light 
 mass group of cosmic rays by correcting for the trigger, reconstruction, 
 and selection efficiency of H and He primaries.

  Before selecting the light subsample, first we apply a set of quality cuts
 on the data.  In particu\-lar, we use similar selection cuts to those described 
 in the previous section, but in order to further reduce the uncertainty in the
 core position, we employed $N_{r40} \geq 60$. In addition, we also remove low 
 energy events ($E_{\textnormal{reco}} < 10^{3.5} \, \mbox{GeV}$ ) from the low efficiency region using $f_{hit} \geq 0.3$, where
 $f_{hit}$ is the fraction of hit PMT's at HAWC during the event \cite{HAWCcrab17}.
 These cuts were applied on both the data and the MC simulations. From
 the latter, we found that the mean core bias of EAS with  
 $E_{\textnormal{reco}} = 10^{4} -  10^5 \, \mbox{GeV}$ is smaller than $11 \, \mbox{m}$, while the 
 energy resolution is $\leq 42 \%$. Here, $E_{\textnormal{reco}}$ is the reconstructed energy, which 
 was estimated using the method 
 described in section 2. We have restricted the analysis to energies 
 $E_{\textnormal{reco}} \leq 10^5 \, \mbox{GeV}$ as the number of hit PMT's is saturated above 
 this threshold energy. The latter can be seen in fig.~\ref{f1}, left, where
 $f_{hit}$ is shown as a function of $E_{\textnormal{reco}}$ for our nominal composition model 
 and for EAS with $\theta < 16.71^\circ$. As we can see on this plot, for 
 vertical cosmic-ray showers, we are constrained to the TeV energy regime. 
 This problem will be solved with the upgrade of HAWC \cite{outtrigger}.
 Meanwhile, we are also investigating the possibility of using 
 inclined EAS to study the high energy regime, since we have observed that
 the threshold energy for saturation increases with  the zenith angle.

   Now, in order to select our light subsample from the selected events, we 
 make use of the age parameter, which is sensitive to the primary composition. 
 This is illustrated, for example, in fig.~\ref{f1}, right, where the mean value 
 of $s$ is presented as a function of $f_{hit}$, for different CR nuclei and 
 vertical events. The plot was derived using MC data for different primaries
 and our nominal model. It shows that the age parameter increases with the 
 mass of the primary nuclei and, in addition, decreases with $E_{\textnormal{reco}}$. 
 This is understood from the fact that light primaries and high energy cosmic rays 
 interact deeper in the atmosphere and hence produce EAS with steeper 
 lateral distributions at ground level. To perform the selection, we apply
 a cut, $s_{He-C}$, located between the predicted curves for He and C (see 
 fig.~\ref{f1}, right) on the data. If the events satisfy $s < s_{He-C}$, 
 then they are classified into the light mass group of cosmic rays,
 otherwise they are considered as a part of the heavy component.  In general,
 according to MC simulations, with our nominal composition model, the retention 
 fraction of protons and helium nuclei in the light subsample is  $\gtrsim 69$\%, while its 
 purity is $\gtrsim 84 \%$ for vertical EAS with $E_{\textnormal{reco}} \geq 10^{3.5} \, \mbox{GeV}$.
 
 For the mass group separation we have used the shower age, $s$ as a function
 of $f_{hit}$. However, we have found that the separation can be also carried 
 out using $E_{\textnormal{reco}}$ instead of $f_{hit}$ with almost no changes in the final 
 results. This is understood from the fact that $f_{hit}$ is proportional 
 to $E_{\textnormal{reco}}$ below the saturation region (c.f. fig.~\ref{f1}, left) and because the 
 energy dependence of $f_{hit}$ is not affected by the mass of the primary nuclei. 
 One important point that should be stressed is that the dependence of 
 $s$ on $f_{hit}$ for an individual element has little dependence on the 
 spectral index, $\gamma$,  of the corresponding energy spectrum. Using
 MC simulations, we found that a change $\Delta \gamma = \pm 0.2$ in the 
 spectrum of the elemental groups leads to variations within $1 \, \%$ in the 
 corresponding $s(f_{hit})$ curves, which implies an independence of
 the $s_{He-C}$ cut from $\gamma$ and the composition model. However, it is
 expected a dependence on the hadronic interaction model, which will be 
 investigated in a further study.

\begin{figure}[t]
     \centering
     \footnotesize
     \includegraphics[width=80mm]{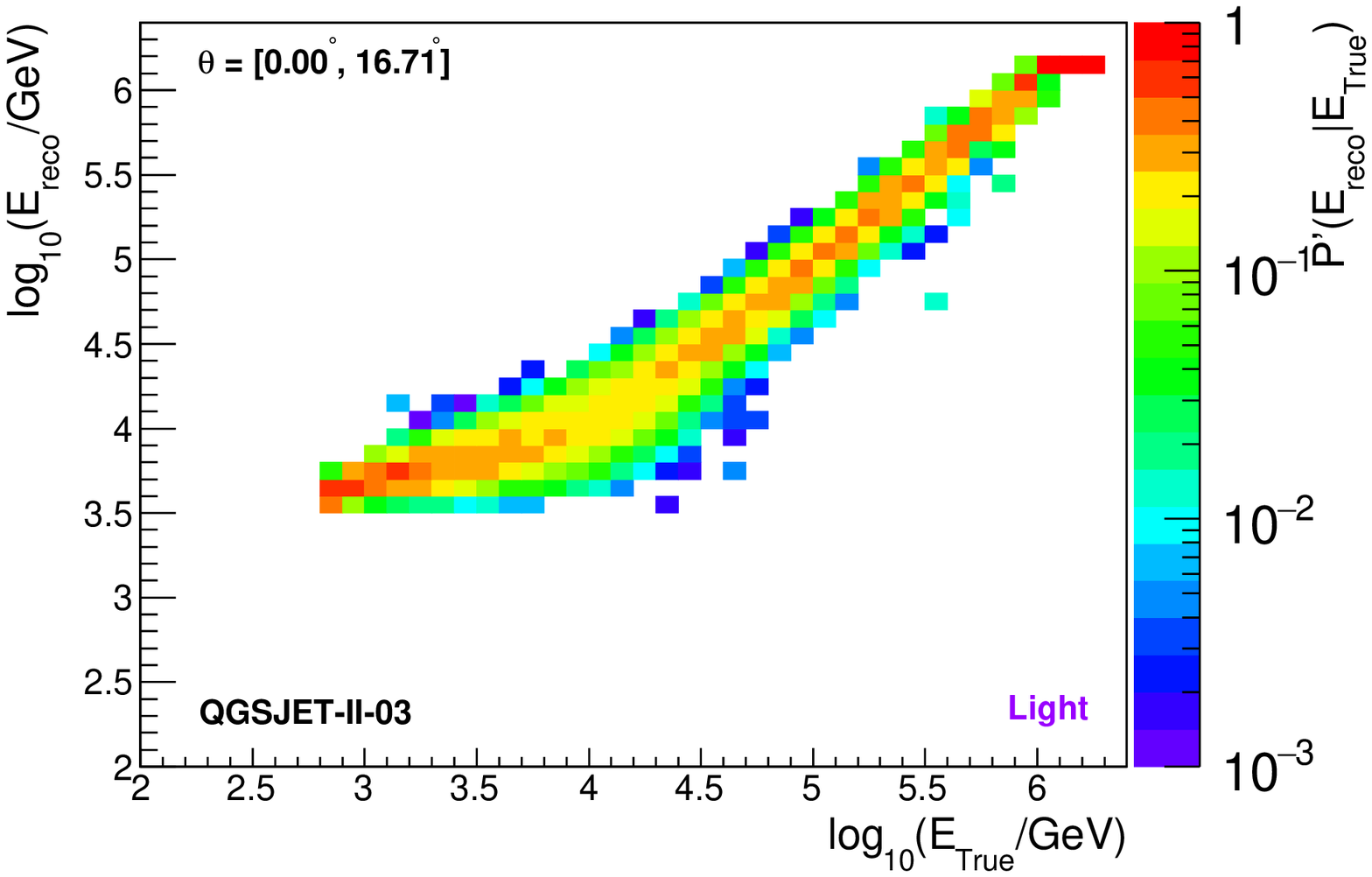}
     \includegraphics[width=65mm]{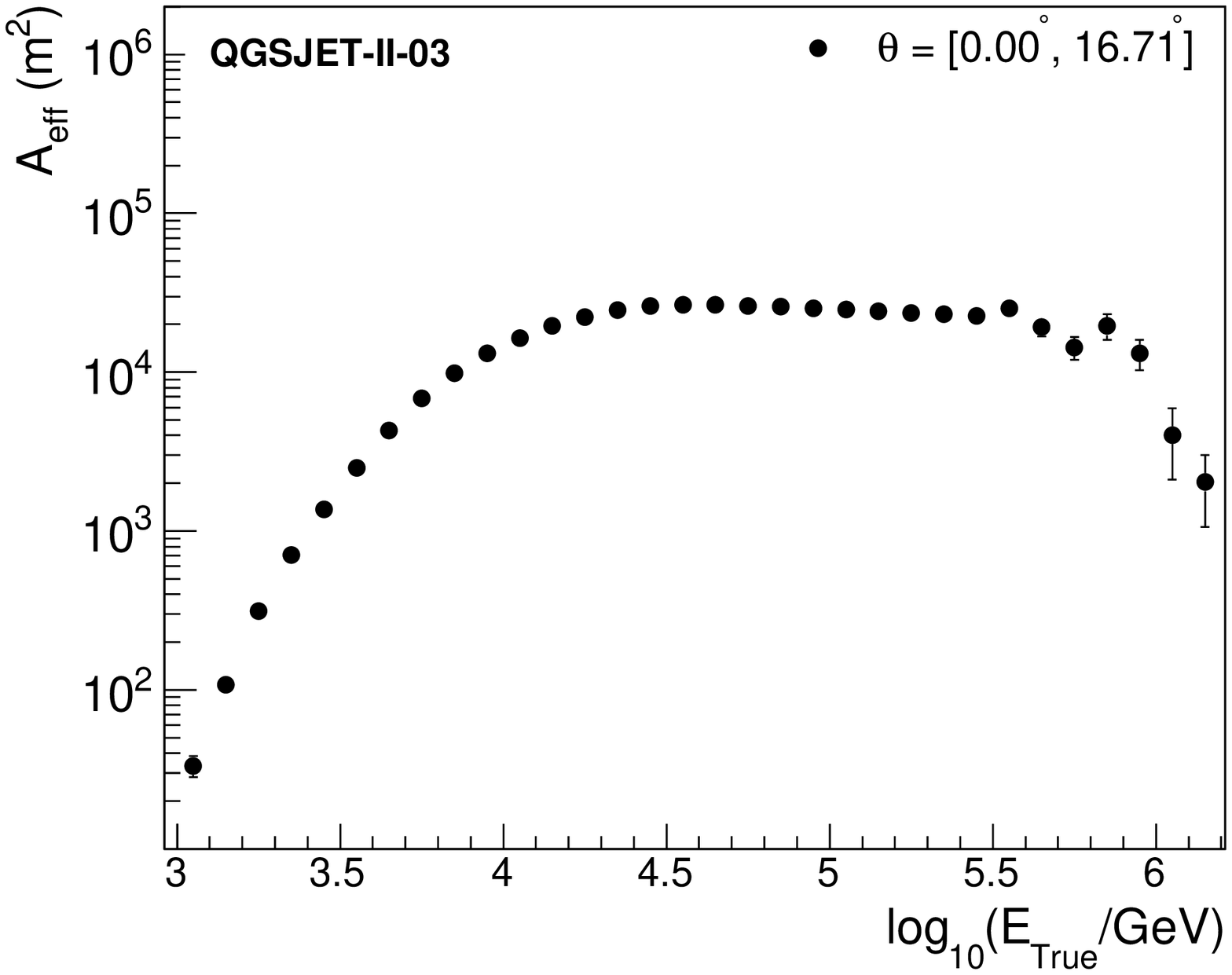} 

     \caption{ \textit{Left panel}: The energy response matrix  for the light subsample
            calculated from  our nominal composition model.
            The vertical axis represent the reconstructed energy, and the horizontal
            axis, the true EAS energy. The color code indicates the probability
            $P^\prime(E_{\textnormal{reco}}|E_{\textnormal{true}})$, which just takes into
            account migration effects and does not considers neither trigger,
            reconstruction  or selection  efficiencies nor corrections from the expected
            fraction of heavy elements in the selected sample. 
            \textit{Right panel}: Effective areas used for the calculation of the 
     		spectra for the light mass group of cosmic rays. The curve was derived from
     		MC simulations using our nominal composition model. At high energies 
            $E_{\textnormal{true}}> 10^{5.5} \, \mbox{GeV}$
            a loss of efficiency is observed in the above effective area due to the 
            saturation effect in $f_{hit}$.
    }
  	\label{f2}
\end{figure}

  After selecting the light subsample, we then build the corresponding energy 
 histogram and apply to it a Bayesian-inspired unfolding procedure \cite{agostini} 
 to correct for migration effects using a response matrix, 
 $P^\prime(E_{\textnormal{reco}}|E_{\textnormal{true}})$, derived from MC simulations 
 for our nominal composition model. Employing the above simulations, we also found 
 some appropriate factors, $f_{corr}$, to be applied to the unfolded spectra to 
 correct it for the expected fraction of heavy elements. These correction factors 
 are defined as the inverse of the expected fraction of H and He primaries inside
 the corresponding energy bin. Finally, the energy spectra 
 for the light component of cosmic rays is estimated by using formula \ref{eq1}. Here, $A_{throw}$ is 
 replaced by  the effective area, $A_{eff}$, which is defined as the product of $f_{corr}$ and the 
 effective area for protons and helium nuclei in the subsample, $A_{eff}^{H+He}$, which 
 is proportional to the efficiency $\epsilon^{H+He}$ for detecting an
 hydrogen/helium-induced EAS and classifying it as a part of our light subsample.
 Both, the response matrix and the effective area are presented in fig.~\ref{f2}. We should emphasize
 that the response matrix for the light sample is defined in such a way that 
 \begin{equation}
 \sum_{j=1} P^\prime(E_{\textnormal{reco}, j}|E_{\textnormal{true}, i}) = 
  \frac{1}{f_{corr}(E_{\textnormal{true}, i}) \cdot \epsilon^{H+He}(E_{\textnormal{true}, i})}\sum_{j=1}  P(E_{\textnormal{reco}, j}|E_{\textnormal{true}, i}) =1,
  \end{equation}
  where $P(E_{\textnormal{reco}}|E_{\textnormal{true}})$ takes into account efficiency effects and it
  is corrected with $f_{corr}$ for the presence of heavy elements in the light subsample.
 
  One of the main sources of systematic uncertainties in the procedure besides
  the influence of the hadronic interaction model is the 
  composition dependence of the response matrix and the effective area. In order
  to evaluate the former, we have estimated both  $P^\prime(E_{\textnormal{reco}}|E_{\textnormal{true}})$ 
  and $A_{eff}$ using different models with distinct elemental abundances as predicted by the 
  Polygonato model \cite{horandel}, and from fits to measurements from 
  ATIC-2 \cite{atic}, MUBEE \cite{mubee} and JACEE \cite{jacee}. 
  
  We will apply the whole procedure above described to the experimental data soon. At the moment,
  we present in fig.~\ref{f1}, right, the  $s(f_{hit})$ distribution of the selected data that we
  will use in the analysis, which consists of events collected during a complete HAWC run 
  performed on June 2nd, 2016. We have employed this particular data set because the 
  MC simulations use the detector configuration of this specific run. The selected showers
  were chosen as described at the beginning of this section but without using the age cut.
  The aforementioned data set contains about $2.3$ million events,
  while the data set with the light subsample contains $\sim 1.4$ million EAS.  
  From fig.~\ref{f1}, right, we see  that the measured distribution follows pretty well 
  the predictions of the nominal composition model, which might indicate a dominance
  of the light component  in the spectrum of cosmic rays in the energy interval 
  $E_{\textnormal{reco}} = 10^{4} -  10^5 \, \mbox{GeV}$. One of the next steps will be 
  to study the robustness of both the method described in this section and the 
  corresponding conclusions with the high-energy hadronic interaction model.

\vspace{3pc}

\noindent
\textbf{Acknowledgements} \\  \\
\small
We acknowledge the support from: the US National Science Foundation 
(NSF); the US Department of Energy Office of High-Energy Physics; the 
Laboratory Directed Research and Development (LDRD) program of Los Alamos 
National Laboratory; Consejo Nacional de Ciencia y Tecnolog\'\i a (CONACyT), 
M\'exico (grants 271051, 232656, 260378, 179588, 239762, 254964, 
271737, 258865, 243290, 132197), Laboratorio Nacional HAWC de 
rayos gamma; L'OREAL Fellowship for Women in Science 2014; 
Red HAWC, M\'exico; DGAPA-UNAM (grants IG100317, IN111315, 
IN111716-3, IA102715, 109916, IA102917); VIEP-BUAP; PIFI 
2012, 2013, PROFOCIE 2014, 2015; the University of Wisconsin Alumni 
Research Foundation; the Institute of Geophysics, Planetary Physics, 
and Signatures at Los Alamos National Laboratory; Polish Science 
Centre grant DEC-2014/13/B/ST9/945; Coordinaci\'on de la Investigaci\'on 
Cient\'\i fica de la Universidad Michoacana. Thanks to Luciano 
D\'\i az and Eduardo Murrieta for technical support.

\end{document}